\documentclass{MSP-template}
\usepackage{ragged2e}
\usepackage{setspace}
\usepackage[noadjust]{cite}
\usepackage{graphicx}
\usepackage{amsmath}
\usepackage{dcolumn}
\usepackage{bm}
\usepackage{float}


\begin{document}

\title{Reconfigurable Three-Dimensional Thermal Dome}

\maketitle

\author{Yuhong Zhou $^a$,}
\author{Fubao Yang $^a$,}
\author{Liujun Xu $^b$,}
\author{Pengfei Zhuang $^a$,}
\author{Dong Wang $^{de}$,}
\author{Xiaoping Ouyang $^{c,*}$,}
\author{Ying Li $^{d,e,*}$,}
\author{Jiping Huang $^{a,*}$}

\begin{affiliations}
$^a$ Department of Physics, State Key Laboratory of Surface Physics, and Key Laboratory of Micro and Nano Photonic Structures (MOE), Fudan University, Shanghai 200438, China

$^b$ Graduate School of China Academy of Engineering Physics, Beijing 100193, China

$^c$ School of Materials Science and Engineering, Xiangtan University, Xiangtan 411105, China

$^d$ Interdisciplinary Center for Quantum Information, State Key Laboratory of Modern Optical Instrumentation, ZJU-Hangzhou Global Scientific and Technological Innovation Center, Zhejiang University, Hangzhou 310027, China

$^e$ International Joint Innovation Center, Key Lab. of Advanced Micro/Nano Electronic Devices \& Smart Systems of Zhejiang, The Electromagnetics Academy of Zhejiang University, Zhejiang University, Haining 314400, China

$^*$ Corresponding author. E-mail: oyxp2003@aliyun.com (X. Ouyang); eleying@zju.edu.cn (Y. Li); jphuang@fudan.edu.cn (J. Huang)

\end{affiliations}

\begin{abstract}

\textbf{ABSTRACT} Thermal metamaterial represents a groundbreaking approach to control heat conduction, and, as a crucial component, thermal invisibility is of utmost importance for heat management. Despite the flourishing development of thermal invisibility schemes, they still face two limitations in practical applications. First, objects are typically completely enclosed in traditional cloaks, making them difficult to use and unsuitable for objects with heat sources. Second, although some theoretical proposals have been put forth to change the thermal conductivity of materials to achieve dynamic invisibility, their designs are complex and rigid, making them unsuitable for large-scale use in real three-dimensional spaces. Here, we propose a concept of a thermal dome to achieve three-dimensional invisibility. Our scheme includes an open functional area, greatly enhancing its usability and applicability. It features a reconfigurable structure, constructed with simple isotropic natural materials, making it suitable for dynamic requirements. The performance of our reconfigurable thermal dome has been confirmed through simulations and experiments, consistent with the theory. The introduction of this concept can greatly advance the development of thermal invisibility technology from theory to engineering and provide inspiration for other physical domains, such as direct current electric fields and magnetic fields.

\end{abstract}


\section{Introduction}
The urgent necessity of rendering objects invisible to infrared detection has sparked significant research into thermal cloaking \cite{APL2008,APL2008-1,LiNC2018,HJ-CPL2020,PR2021,HJ-IJHMT2021,LiNC2022,Huang20,Yeung22,LIAM2023,JPPNAS23,NM2012,HuEPL15,FujiiIJHMT19,QinIJHMT19,MTP2022}. Traditionally, thermal invisibility cloaks were designed by initially enveloping the target with insulating materials, followed by guiding the heat flow around the cloaked area, thus achieving invisibility. Various theories such as transformation-thermotics \cite{APL2008,APL2008-1,PREDai,HJ-PRE2018,HJ-JAP2018,AFM2020,HJ-PRAP2020,HJ-ES2020,AM2021,NRM2021}, scattering-cancellation \cite{PRL2014.1,PRL2014.2,PRL2014.3,HanAM18,PRAP2022}, and topology-optimization \cite{FujiiAPL18,ShaNC21,JiIJHMT22,HirIJHMT22,2022MTP2}have been developed following this approach.

However, the thermal invisibility devices designed based on the above principles are facing significant engineering challenges, which contrasts with the tremendous potential for other metamaterials in engineering applications \cite{HJ-ESEE2019,PER2021,E2022-1,E2022-2,E2023}. They pose manufacturing and installation difficulties, problems with reusability, and most critically, completely enclosed designs cannot accommodate internal heat sources. This is because the continuous rise in internal temperature could result in disastrous outcomes \cite{SP2020}. It's noteworthy that scenarios requiring the concealment of heat-generating objects are quite common, yet have been conspicuously overlooked in previous studies. To overcome these limitations, researchers have begun exploring non-traditional cloak \cite{GaoEPL13,JAP2019}designs that do not require the cloaked area to be fully enclosed, thereby allowing interaction with the external environment. However, such studies have primarily focused on two-dimensional structures and have relied heavily on negative thermal conductivity in their designs, a feature that poses considerable challenges in real-world applications. Moreover, these newly proposed ``external cloaks'' fail to address a critical problem - the dissipation of internal heat sources, which is an essential requirement for many practical applications. Hence, there's a clear need for further research to address these issues, potentially opening new avenues in the field of thermal cloaking.

Furthermore, reconfigurable capabilities significantly amplify a device's potential to adjust to dynamic requirements, thereby boosting its applicability and acceptance across various engineering sectors. Existing thermal cloaking devices, however, are usually tailored to a specific background; changes in the environment necessitate a redesigned cloak, making the process inconvenient and economically inefficient. Indeed, researchers have proposed numerous methods, such as nonlinearity~\cite{PRL2015}, chameleon-like behaviors~\cite{PRAP2019,PRAP2020}, convection~\cite{NM2019,LiAM20,NC2020}, and height manipulation~\cite{AM2022,IJHMT2023}, to regulate the thermal conductivity of materials, aiming to meet dynamic stealth requirements. However, translating these techniques into three-dimensional spaces presents significant hurdles, rendering them impractical for many real-world applications.

\begin{figure}[htb]
	\centering
	\includegraphics[width=\linewidth]{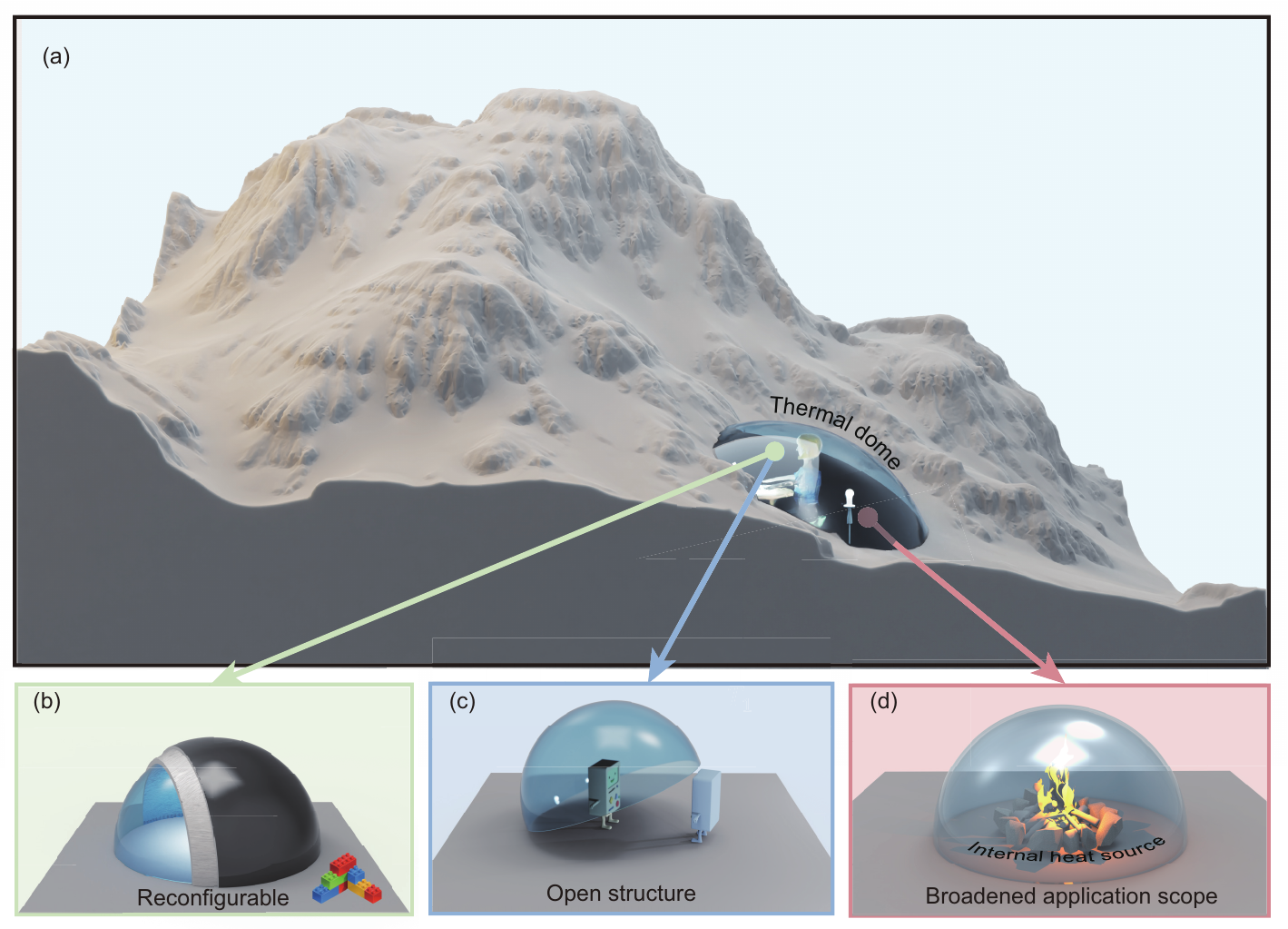}\\
	\caption{(a) displays the application scenario of the thermal dome, which is able to shield the target from infrared detection. (b), (c), and (d) highlight the unique features and advantages of the thermal dome compared to traditional thermal cloaks. Specifically, (b) shows the reconfigurable nature of the thermal dome, enabling it to adapt to changing environments. (c) demonstrates the open structure of the thermal dome, which makes it extremely convenient in practical use, such as the replacement of the protected target as shown in the figure. (d) illustrates that the thermal dome remains applicable even when the target object generates heat, thereby expanding its range of applications.}
	\label{F1}
\end{figure}

This paper introduces a groundbreaking solution to these challenges: the thermal dome (see Fig.~\ref{F1}). A device uniquely designed with practical applications in mind, the thermal dome showcases an open hidden area, facilitating easy installation and reuse. Remarkably, this device accomplishes thermal invisibility for heat-generating objects—a pioneering achievement in the field. Inspired by Lego structures ~\cite{LSA2022,SA2021}, we have fused an open architecture with a multi-layered design, granting the thermal dome a reconfigurable nature. Users can intuitively assemble it to meet specific requirements and adapt to varying environments, much like assembling Lego blocks. This level of flexibility and adaptability sets the thermal dome apart from conventional thermal cloaks, underscoring its exceptional engineering significance. Through the resolution of differential equations, we designed a semi-ellipsoidal thermal dome and validated its functionality in a hemispherical form using common bulk materials. The introduction of this thermal dome concept marks a paradigm shift in the field of thermal invisibility devices, propelling them into a phase of practicality and inspiring further exploration of their feasibility in real-world scenarios.

\section{Design principles of thermal domes}

\begin{figure}[htb]
	\centering
	\includegraphics[width=\linewidth]{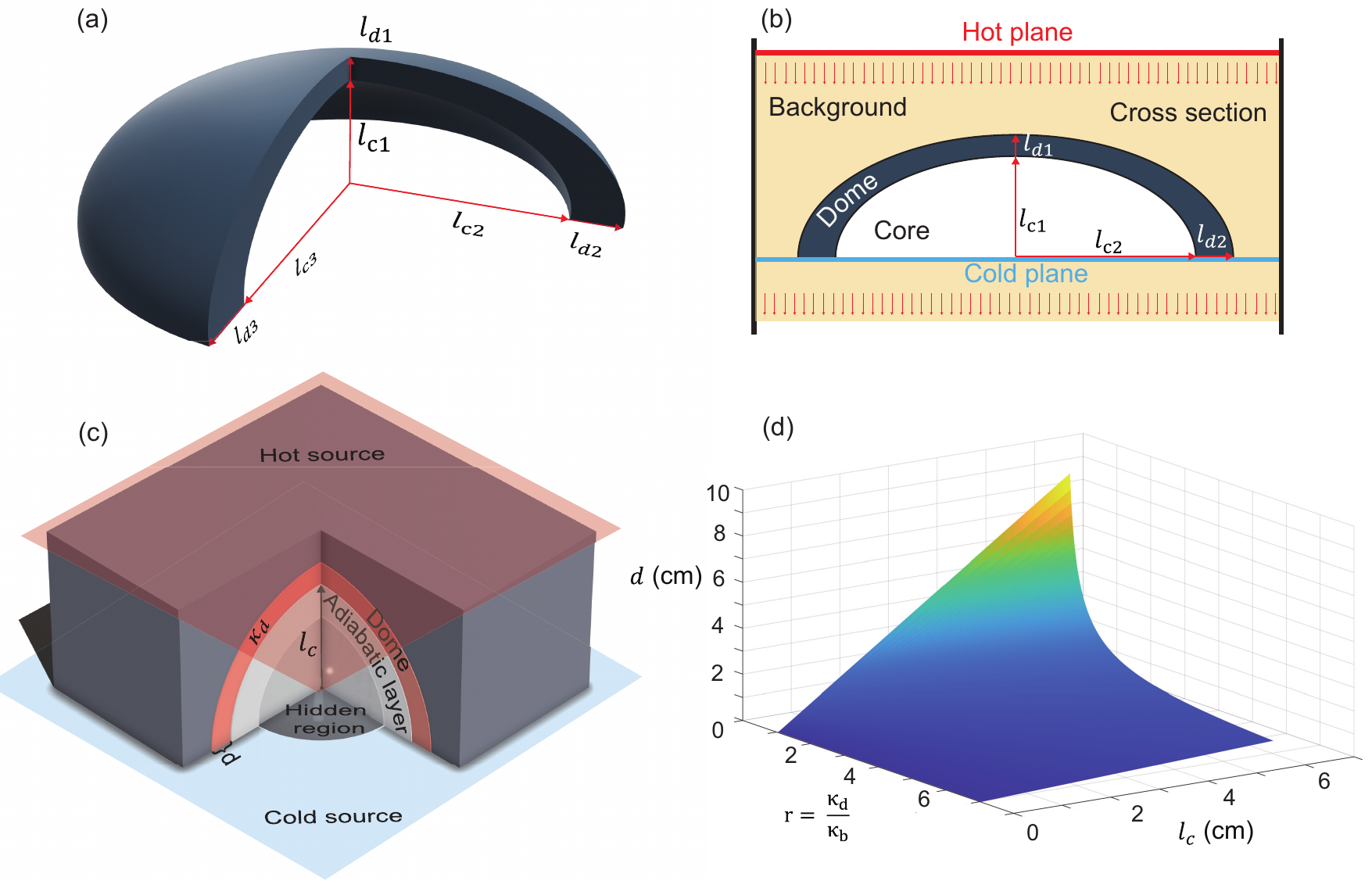}
	\caption{(a) presents a schematic representation of the thermal dome, while (b) exhibits its cross-sectional view. (c) presents a cross-sectional view of a single-layer hemispherical thermal dome. In (b), we show the thickness of a single-layer hemispherical thermal dome as a function of $l_{\rm c}$ and $r$.}
	\label{F2}
\end{figure}

We consider three-dimensional heat transfer in the absence of convection and radiation. In a homogeneous medium with a constant thermal conductivity of $\kappa_{\rm b}$, heat flows uniformly from the high-temperature surface towards the low-temperature surface. However, the introduction of a target object disturbs the heat flow due to the object's different thermal conductivity, $\kappa_{\rm c}$. To eliminate this disturbance, a thermal dome can be placed on the target, which cloaks the target as an object with the same thermal conductivity as the background, achieving the purpose of stealth. To achieve this functionality, the shape and material of the thermal dome must be carefully designed. While the shape of the thermal dome can be of any form, choosing a shape with poor symmetry can result in irregular faces of the thermal dome, making the design process challenging. The semi-ellipsoidal shape has excellent symmetry and can form a variety of shapes by changing the lengths of its three axes to meet the designer's needs, making it an ideal choice for the shape of a thermal dome. As shown in Fig.~\ref{F2}(a), the semi-axis of the core (dome) is specified as $l_{ci}$ ($l_{di}$) along the $x_i$ axis, where $i=1,2,3$ represents the three dimensions. The heat transfer equation in the ellipsoidal coordinate system is written as\cite{Milton02}
\begin{equation}\label{e1}
	\frac{\partial}{\partial\rho_1}\left[g(\rho_1)\frac{\partial T}{\partial\rho_1}\right] + \frac{g(\rho_1)}{\rho_1 + l_{i}^{2}}\frac{\partial T}{\partial\rho_1} = 0,
\end{equation}
where $g\left(\rho_1\right)=\prod\limits_i\left(\rho_1+l_{i}^2\right)^{1/2}$. We implement an external thermal field along the $x_i$ axis as depicted in Fig.~\ref{F2}b. The semi-ellipsoid is visualized as a complete ellipsoid halved, with the solution procedure identical to that of the complete ellipsoid barring an additional boundary condition: the entire surface temperature where the base of the dome is located must equalize to ensure the background temperature field remains undisturbed. In addition, to solve the differential equations, we need to impose some additional boundary conditions, namely, the equality of temperatures and normal heat fluxes at the interfaces between different regions. Applying the generalized solution to the boundary conditions, we obtain the design requirement for the thermal dome:
\begin{equation}\label{e1}
	\kappa_{b}=\frac{L_{ci}\kappa_{c}+\left(1-L_{ci}\right)\kappa_{d}+\left(1-L_{di}\right)\left(\kappa_{c}-\kappa_{d}\right)f}{L_{ci}\kappa_{c}+\left(1-L_{ci}\right)\kappa_{d}-L_{di}\left(\kappa_{c}-\kappa_{d}\right)f}\kappa_{d},
\end{equation}
where $f=g\left(\rho_{c}\right)/g\left(\rho_{d}\right)=\prod\limits_il_{ci}/l_{di}$ denotes the volume fraction, $L_{ci}$ and $L_{di}$ are the shape factors. The detailed solution steps are outlined in Supplementary Note 1.

As previously mentioned, besides the thermal conductivity condition, another requisite for the semi-ellipsoidal structure not to impact the background temperature field is to have the entire surface on which its base is located at an equal temperature. Furthermore, if the thermal conductivity of the thermal dome needs to be independent of the core region, the core region can be insulated, whereby $\kappa_{c}$ can be considered as equal to 0Wm$^{-1}$~K$^{-1}$, making the designed thermal dome applicable to arbitrary objects. Specifically, for a hemispherical thermal dome, as shown in Fig.~\ref{F2}(c), we can establish its thickness in relation to its geometric size and thermal conductivity
\begin{equation}\label{e3}
	d=\left(\sqrt[3]{\frac{2r+1}{2r-2}}-1\right)l_{\rm c}.
\end{equation}
$r={\kappa_{\rm d}}/{\kappa_{\rm b}}$ represents the ratio of the thermal conductivity of the thermal dome to that of the background. Importantly, at this stage, we have introduced an insulating layer beneath the hemispherical thermal dome to ensure its functionality across various objects.

The inner diameter $l_{\rm c}$ of the thermal dome is determined by the specific usage scenario, while the thickness of the dome is closely related to the choice of material, as seen from Eq.~(\ref{e3}). The thickness variation of a single-layer thermal dome as a function of $l_{\rm c}$ and $r$ is depicted in Fig.~\ref{F2}(d). The ratio of $r$ has a significant impact on the thickness $d$, with $d$ becoming infinite when the device uses the same material as the background (i.e., $r=1$). Conversely, when the thermal conductivity of the thermal dome $\kappa_{\rm 1}$ is much larger than that of the background $\kappa_{\rm b}$ (i.e., $r$ is a large number), the thickness of the thermal dome becomes very small. This is because the thermal dome acts as a compensation for thermal conductivity, and if its material has a high thermal conductivity, only a small amount of material is needed to compensate. Therefore, selecting an appropriate material to manufacture the thermal dome based on the specific scenario is crucial. For instance, if the background is made of cement and a thin thermal dome is preferred, copper would be a better choice. According to Eq.~(\ref{e3}), the thickness of the layer decreases to 0.16 mm when $l_{\rm c}$ is set to 10 cm in this case.

The aforementioned approach can be readily expanded to accommodate core-shell structures with $n$ layered shells. We can employ computational software to determine the parameters for the thermal domes in each layer. An expedited method for designing a multilayer thermal dome involves leveraging effective medium theory, where the design process can be conducted iteratively, layer by layer. Further details are provided in Supplementary Note 2.

\section{Function verification and simulation results}

We utilized the commercial software COMSOL Multiphysics to execute finite-element simulations and authenticate our theoretical design. We carried out steady-state simulations with the Heat Transfer Module, and the transient results will be elaborated in Supplementary Note 3. For simplicity, we use the hemispherical thermal dome for verification. The background is dimensioned at 30 $\times$ 30 $\times$ 15 cm$^3$ and features a thermal conductivity, $\kappa_{\rm b}$, of 10 W~m$^{-1}$~K$^{-1}$. An object represented by a core with $R_o$=9cm and $\kappa_{\rm o}$=500 W~m$^{-1}$~K$^{-1}$ is introduced. The temperature and isotherm distributions in the background illustrate the perturbation of heat flow. Once heat transfer reaches equilibrium, we analyze the efficacy of our thermal dome through the examination of the temperature distribution in three distinct groups.

\begin{figure}[htbp]
\centering\includegraphics[width=0.8\linewidth]{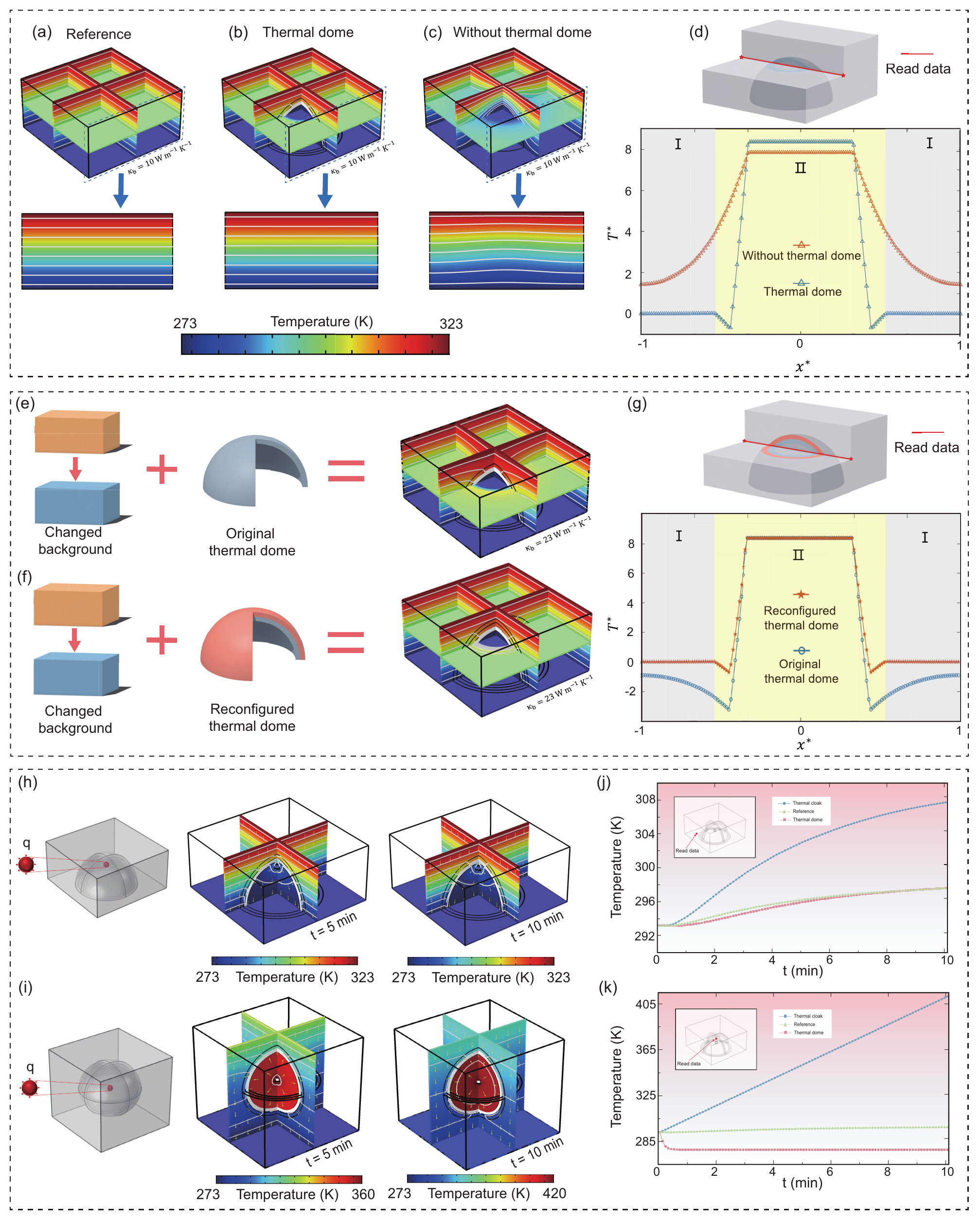}
\caption{Temperature distributions for different groups. (a)-(c) shows the temperature distributions of the reference, with a thermal dome and without a thermal dome, respectively. (d) shows imensionless temperature on a chosen line. Region one represents the background, and region two represents the thermal dome and the object. (e) and (f) demonstrate the reconfigurable capability of the thermal dome to adapt to changing backgrounds. (e) shows the simulation result for the original thermal dome in a changed background, (f) shows the simulation result for the reconfigured thermal dome in a changed background. In this case, we added a new layer ($\kappa_{\rm 2}$=85~W~m$^{-1}$~K$^{-1}$, $d$=1~cm) to the original thermal dome. (g) shows dimensionless temperature on a chosen line. (h) and (i) compare the performance of the thermal dome and thermal cloak with a heat source in the hidden area at 5 minutes and 10 minutes respectively. The heat source in the thermal dome and thermal cloak emitting heat outward at a rate of 500 kW per square meter. In (j) and (k), we quantitatively show the variation of temperature at different positions with time. }
\label{F3}
\end{figure}

As depicted in Fig.~\ref{F3}(c), the presence of the object distorts the temperature distribution of the background, causing the isotherms to bend and reflect the disturbance. In contrast, the background devoid of the object exhibits a uniform temperature distribution and straight isotherms, as shown in Fig.~\ref{F3}(a). Upon placement of the thermal dome over the object (Fig.~\ref{F3}(b), featuring a shell with an inner diameter $R_{\rm o}$=9 cm, thickness $d$=1 cm, and $\kappa_{\rm a}$=0.023 W~m$^{-1}$~K$^{-1}$ representing the adiabatic layer, and another shell with inner diameter $R_{\rm 1}$=10 cm, thickness $d$=1 cm, and $\kappa_{\rm 1}$=55 W~m$^{-1}$~K$^{-1}$ representing the thermal dome), the temperature distribution and isotherms in the background precisely match those of the reference group.

To accurately contrast their differences, we exported data from a cross-section at $z$=7.5 cm (-15 cm $< x <$ 15~cm) (Fig.~\ref{F3}(d)). We utilized a dimensionless temperature $T^{*}=100(T_{\rm0}-T)/T_{\rm0} $ and a dimensionless position of plot $x^{*}=2x/L $, where $T_{\rm0}$ and $L$ denote the temperature of the reference and the length of the background, respectively. The temperature without the thermal dome (depicted by an orange line) diverges from the temperature of the reference across the entire space. Conversely, the blue line (with the thermal dome) aligns perfectly with $T_{\rm0}$ in the background, signifying successful achievement of the thermal cloaking effect by the thermal dome.

The scenario of a reconfigured thermal dome is considered when the background thermal conductivity alters. In such a case, with $\kappa_{\rm b}$ changing from 10 W~m$^{-1}$~K$^{-1}$ to 23 W~m$^{-1}$~K$^{-1}$, the original thermal dome no longer satisfies the stealth requirement (Fig.~\ref{F3}(e)). One solution is to append another layer to the outer surface of the single-layer thermal dome (Fig.~\ref{F3}(f)). The Lego-like structure of the thermal dome facilitates easy layering or removal, a feat challenging in traditional thermal cloaks. Upon reconfiguration, the new thermal dome exhibits impressive performance in the new background. To verify the preceding discussion, we plot the dimensionless temperature $T^{*}$ (Fig.~\ref{F3}(g)).

Past designs of thermal invisibility cloaks failed to account for scenarios with heat sources within the concealed area. However, a significant number of real-world objects to be hidden do emit heat, rendering conventional thermal cloaks ineffective. The open structure of the thermal dome allows the hidden area to directly engage a cold source that absorbs the heat generated within the hidden area. This ensures that the invisibility function remains uncompromised, and the internal temperature does not continually elevate. We showcase the simulation validation of the thermal dome and the conventional thermal cloak in Fig.~\ref{F3}(h) and Fig.~\ref{F3}(i), respectively.

As the time extends, the temperature within the thermal dome remains virtually stable, whereas the temperature within the traditional thermal cloak's space continues to rise. For further scrutiny, we selected two points inside the hidden region and in the background area, and plotted their temperature values over time in Fig.~\ref{F3}(j) and Fig.~\ref{F3}(k). The comparison with the pure background reference group reveals the disadvantages of traditional thermal cloaks when heat sources are present in the hidden area: first, the internal temperature incessantly escalates over time; second, in the absence of entirely insulating materials, the elevated internal temperature impacts the temperature distribution in the background, leading to invisibility function failure. In contrast, the thermal dome not only preserves excellent invisibility but also maintains a stable internal temperature, akin to the cold source temperature. Thus, the thermal dome emerges as an effective solution for concealing objects with heat sources.

Throughout the preceding discussion, regardless of whether we consider a single-layer or multi-layer thermal dome, the temperature bias is vertical, directing heat flow from top to bottom. However, altering the direction of the temperature bias to horizontal does not compromise the perfect cloaking function of the thermal dome, as shown in Supplementary Note 4. When the temperature bias is arbitrary (Fig.S5), the temperature field becomes non-uniform, and the cloaking function of the thermal dome is no longer perfect. Yet, the simulation results reveal that the thermal dome still offers a substantial amount of cloaking effect compared to the control group without the thermal dome. Under such conditions, the object remains undetectable by low-precision infrared cameras. Furthermore, other shapes of thermal domes will be discussed in Supplementary Note 5.

\section{Experimental validation of the thermal dome}
\subsection{Experimental results}

\begin{figure}[H]
	\includegraphics[width=\linewidth]{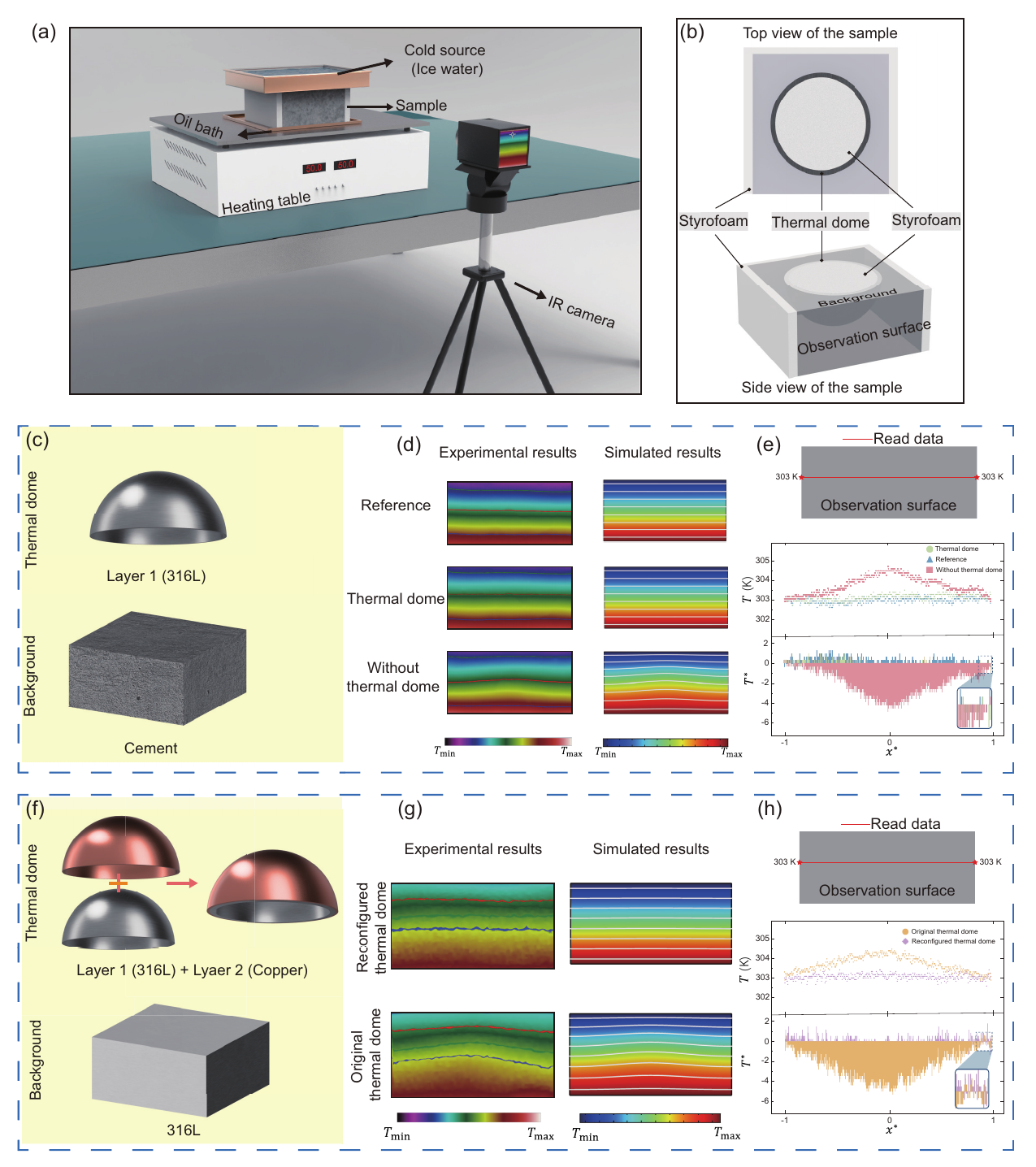}
	\caption{Experimental demonstration of the thermal dome. (a) depicts the schematic diagram of the experimental setup, in which the sample is positioned between the cold (273 K) and hot (323 K) sources, with the open side connected to the cold source. (b) shows the structure of the sample, which is filled with foam to prevent thermal convection and acts as both the adiabatic layer and object. (c) illustrates the materials used for the single-layer thermal dome and the background. The experimental results with corresponding simulation results for three groups of samples are presented in (d). In (e), temperature data from a single line in three sets of samples were extracted to further compare their similarities and differences. (f) displays the materials used for the new multi-layer thermal dome and the new background. (g) shows the experimental results with corresponding simulation results for the group with new multi-layer thermal dome and the group with the original single-layer thermal dome. Similarly, in (h), temperature differences between the two were quantitatively analyzed at the same location.}
	\label{F4}
\end{figure}

We performed experimental validation of the hemispherical thermal dome and the results are depicted in Fig.~\ref{F4}. Constraints of the experimental setup meant that the heat source was situated below, and the cold source was positioned above. The temperature distribution within the background can be inferred from its surface. Any disturbance in the background heat flow will lead to distortion in its surface isotherms; otherwise, the isotherms remain straight. Therefore, we employed an infrared camera to assess the temperature distribution on the sample's surface, thus validating the function of the thermal dome.

In the single-layer thermal dome experiment (Fig.~\ref{F4}(c)), Layer 1 was utilized as the thermal dome, while cement served as the background. Three samples were prepared: a reference group comprised solely of the background, a control group with the presence of objects but without a thermal dome, and a group with objects safeguarded by the thermal dome. The experimental and simulation results for these three groups are presented in Fig.~\ref{F4}(d). Both the experimental and simulation outcomes demonstrate that the isotherms of the group without a thermal dome are distorted, whereas the isotherms of the group equipped with a thermal dome, and the reference group, are straight. This signifies that the thermal dome shields the object, preventing its detection. For an intuitive comparison, data from the same line were scrutinized, as shown in Fig.~\ref{F4}(e). Here, $T$ denotes the temperature, and $T^{*}=100(303-T)/303$ is a dimensionless temperature that can represent deviation. Both the reference group and the group with a thermal dome exhibit a temperature close to 303~K, with deviations approximating 0 K. The group without a thermal dome shows a significantly larger temperature deviation.

Fig.~\ref{F4}(f) delineates the functionality of the reconfigured thermal dome in a new background, where stainless steel (316L) is used as the background material, and the thermal dome comprises Layer 1 and Layer 2. Experimental results with their corresponding simulation results for the group with the new thermal dome and the group with the original thermal dome are displayed in Fig.~\ref{F4}(g). Additionally, $T$ and $T^{*}$ are plotted in Fig.~\ref{F4}h to visually compare the efficacy of the new multi-layer thermal dome with the original single-layer dome. From Fig.~\ref{F4}g and Fig.~\ref{F4}(h), we can affirm the following: multi-layer thermal domes, with their Lego-like structures, can adapt to various environments by simply adjusting the number of layers – a feat challenging to achieve with traditional thermal cloaks.

In real-world applications, there are inevitable additional factors that affect the functionality of the thermal dome. For instance, during the assembly of a multi-layer thermal dome, the thermal contact resistance (TCR) between different layers can have an impact. In Supplementary Note 6, we discuss various factors influencing the TCR and simulate the temperature distribution of the thermal dome considering thermal contact resistance. We also explore strategies to mitigate the influence of TCR on the functionality of the thermal dome. Additionally, convective and radiative heat transfer between the sample and the external environment cannot be disregarded in practical applications. In Supplementary Note 7, we address these factors and conclude that the thermal dome maintains its functionality when convective and radiative heat transfers do not significantly alter the original temperature distribution – that is, when the temperature field remains relatively uniform.

\subsection{Experimental setups}
\textbf{Background Materials}

Cement: Dimensions: 15 $\times$ 15 $\times$ 7.5 cm, Thermal conductivity: $\kappa_{\rm b}$ = 1.28 W m$^{-1}$ K$^{-1}$. The cement used in this study is a common construction material. Its thermal conductivity varies depending upon moisture content and curing time. We maintained uniform thermal conductivity across all sample sets by using cement from the same batch and initiating the curing process simultaneously.

316L Steel: Dimensions: 15 $\times$ 15 $\times$ 7.5 cm, Thermal conductivity: $\kappa_{\rm b}$ = 16.2 W m$^{-1}$ K$^{-1}$. This steel variant was shaped using computer numerical control (CNC) machining.

\textbf{Conductive Layers}

Layer 1: A 316 stainless steel layer created using 3D printing, with an inner diameter $R_1$ of 6 cm, thickness $d$ of 0.25 cm, and thermal conductivity $\kappa_1$ of 16.2 W m$^{-1}$ K$^{-1}$.

Layer 2: A copper layer created through CNC machining, having an inner diameter $R_1$ of 6.25 cm, thickness $d$ of 0.12 cm, and thermal conductivity $\kappa_2$ of 385 W m$^{-1}$ K$^{-1}$.

\textbf{Target Object and Adiabatic Layer}

Constructed of styrofoam for simplicity, both the target object and the adiabatic layer possess the same thermal conductivity, $\kappa_{\rm o}$ = $\kappa_{\rm a}$ = 0.042 W m$^{-1}$ K$^{-1}$, and a maximum radius $R$ of 6 cm.

\textbf{Thermal Sources}

Hot Source: This is comprised of a heating table set to a constant temperature of 323~K and a copper oil bath pan, which is placed directly on the heating table and filled with silicone oil.

Cold Source: A copper pan filled with an ice-water mixture acts as a cold source, maintaining a temperature of 273~K.

\textbf{Sample Preparation}

Three sets of samples were prepared: a reference group, a thermal dome group, and a group without a thermal dome. We began by fabricating three 15 $\times$ 15 $\times$ 7.5 cm cuboid molds using polyvinyl chloride foam boards, followed by a common batch of cement. The reference group involved simply pouring the cement into the mold, followed by mixing and drying. For the thermal dome group, we fixed the thermal dome to the mold bottom, poured cement, mixed it, and allowed it to dry. The group without a thermal dome involved a similar process, but with a resin shell manufactured via 3D printing that matched the thermal dome's size. After a fortnight, the cement was fully dried, the molds and resin shell removed, and the resulting voids filled with styrofoam. For samples featuring a 316L steel background, we directly placed the thermal dome into the pre-formed background.

\textbf{Experimental Setup Construction}

To minimize convective heat transfer between the sample surface and the environment, we first wrapped two layers of foam around the sample. We then prepared a copper basin for the oil bath, which was placed on a heating platform set to 323 K, filled with silicone oil. This oil bath heating method ensured consistent heating of the bottom surface, eliminating inconsistencies caused by gaps between the sample and the heating platform. To serve as a cold source, another copper basin filled with an ice-water mixture was placed atop the sample, with the intervening gap filled with silicone grease to ensure effective contact. To avoid uneven temperatures caused by disparate ice distribution, we maintained the ice in a single block floating atop the water. The ambient temperature in the laboratory was kept at 298 K.

\textbf{Data Collection}

Post-heating, we collected temperature data from the observation surface using an infrared camera at ten-second intervals. To mitigate environmental interference, we used a blackboard as a backdrop behind the sample. We also applied a transparent film on the sample surface and used a black foam backdrop to reduce the effect of material emissivity on temperature measurements. While these measures may not entirely eliminate emissivity effects, they ensure accurate temperature differences between samples under identical environmental conditions, providing reliable conclusions.

\section{Conclusion and discussion}

In this paper, we deviate from the traditional approach of designing thermal cloaking devices and propose an entirely new concept for achieving thermal invisibility - directing the heat flow directly towards an isothermal surface. The structure of the cloaking device under this notion is no longer closed, but open, allowing its internal hidden space to directly interact with the outside world. We evocatively refer to such a cloaking device as the ``thermal dome''. Compared to traditional thermal cloaks, the open structure of this device makes it more versatile in various scenarios, including those with internal heat sources, while also simplifying its reuse in practical applications. The combination of multilayered and open structures allows the thermal dome to be assembled like Lego, providing it with reconfigurability. This property enables users to customize the thermal dome to specific environmental conditions, greatly improving its practicality. Moreover, we have manufactured the hemispherical thermal dome using common materials, eliminating the need for extreme materials required by some other thermal cloaks. Both simulation and experimental results confirm the functionality of the thermal dome. 

It's worth noting that while we have set the base of the thermal dome to be an isothermal boundary condition in our paper to achieve zero disturbance to the background, it is entirely feasible to replace this condition with a substantial heat reservoir in reality. Under such circumstances, the disturbance caused by heat flow directed towards the heat reservoir through the thermal dome can be considered negligible. Alternatively, if we merely need to prevent disturbance to the background temperature at a particular location caused by an object, we can effectively use the thermal dome to guide the heat flow passing through the object and disperse it to less crucial locations. This novel approach opens up a new research direction in the field of thermal invisibility. Further exploration of this concept could provide substantial theoretical support for the practical application of thermal cloaking devices. This concept can be extended to other functional thermal devices\cite{HJ-AIPA2015,HJ-APL2016,HJ-APL2017,HJ-PRAP2019,HJ-IJHMT2020,HJ-PRL2022} and even to other fields\cite{HJ-PRE2003,HJ-JAP2004,HJ-APL2005,HJ-JPCC2007,HJ-PA2008,HJ-PO2013,HJ-LPCB2015}.

\medskip
\textbf{Acknowledgements}\par

This work was supported by the National Natural Science Foundation of China to J.H. (12035004), the Science and Technology Commission of Shanghai Municipality to J.H. (20JC1414700), and the National Natural Science Foundation of China to Y.L. (92163123 and 52250191), and the Fundamental Research Funds for the Central Universities to Y.L. (2021FZZX001-19).

\medskip
\noindent\textbf{Compliance with ethics guidelines} \par 
\noindent The authors declare that they have no conflict of interest or financial conflicts to disclose.

\medskip
\noindent\textbf{Appendix A. Supplementary information} \par 
\noindent All data are available in the manuscript or the Supplementary information.


\medskip

\begin{thebibliography}{99}

\bibitem{APL2008} Fan CZ, Gao Y, Huang JP. Shaped graded materials with an apparent negative thermal conductivity. Appl Phys Lett 2008; 92(25): 251907. https://doi.org/10.1063/1.2951600.

\bibitem{APL2008-1} Chen T, Weng CN, Chen JS. Cloak for curvilinearly anisotropic media in conduction. Appl Phys Lett 2008; 93(11): 114103. https://doi.org/10.1063/1.2988181.

\bibitem{NM2012} Zheludev NI, Kivshar YS. From metamaterials to metadevices. Nat Mater 2012; 11(11): 917-924. https://doi.org/10.1038/nmat3431.

\bibitem{LiNC2018} Li Y, Bai X, Yang T, Luo H, Qiu CW. Structured thermal surface for radiative camouflage. Nat Commun 2018; 9: 273. https://doi.org/10.1038/s41467-017-02678-8.

\bibitem{HJ-CPL2020} Xu LJ, Huang JP. Active Thermal Wave Cloak. Chin Phys Lett 2020; 37(12): 120501, https://doi:10.1088/0256-307x/37/12/120501.

\bibitem{PR2021} Yang S, Wang J, Dai GL, Yang FB, Huang JP. Controlling macroscopic heat transfer with thermal metamaterials: Theory, experiment and application. Phys Rep 2021; 908: 1-65. https://doi.org/10.1016/j.physrep.2020.12.006.

\bibitem{LiNC2022} Li Y, Qi M, Li J, Cao PC, Wang D, Zhu XF, et al. Heat transfer control using a thermal analogue of coherent perfect absorption. Nat Commun 2022; 13: 2683. https://doi.org/10.1038/s41467-022-30023-1.

\bibitem{MTP2022} Martinez F, Maldovan M. Metamaterials: Optical, acoustic, elastic, heat, mass, electric, magnetic, and hydrodynamic cloaking, Mater Today Phys 2022; 27: 100819. https://doi.org/10.1016/j.mtphys.2022.100819.

\bibitem{Huang20} Huang JP. Theoretical Thermotics: Transformation Thermotics and Extended Theories for Thermal Metamaterials. Singapore: Springer; 2020.

\bibitem{Yeung22} Yeung WS, Yang RJ. Introduction to Thermal Cloaking: Theory and Analysis in Conduction and Convection. Singapore: Springer; 2022.

\bibitem{HJ-IJHMT2021} Xu LJ, Wang J, Dai GL, Yang S, Yang FB, Wang B, et al. Geometric phase, effective conductivity enhancement, and invisibility cloak in thermal convection-conduction. Int J Heat Mass Transf 2021; 165: 120659. https://doi.org/10.1016/j.ijheatmasstransfer.2020.120659.

\bibitem{JPPNAS23} Jin P, Liu JR, Xu LJ, Wang J, Ouyang XP, Jiang JH, et al. Tunable liquid-solid hybrid thermal metamaterials with a topology transition. Proc Natl Acad Sci USA 2023; 120(3): e2217068120. https://doi.org/10.1073/pnas.2217068120.

\bibitem{LIAM2023} Ju R, Xu GQ, Xu LJ, Qi MH, Wang D, Cao PC, et al. Convective Thermal Metamaterials: Exploring High-Efficiency, Directional, and Wave-Like Heat Transfer. Adv Mater 2023; 35(23): 2209123. https://doi.org/10.1002/adma.202209123.

\bibitem{HuEPL15} Hu R, Xie B, Hu J, Chen Q, Luo X. Carpet thermal cloak realization based on the refraction law of heat flux. Epl 2015; 111(5): 54003. https://doi.org/10.1209/0295-5075/111/54003.

\bibitem{FujiiIJHMT19} Fujii G, Akimoto Y. Topology-optimized thermal carpet cloak expressed by an immersed-boundary level-set method via a covariance matrix adaptation evolution strategy. Int J Heat Mass Transf 2019; 137: 1312-1322. https://doi.org/10.1016/j.ijheatmasstransfer.2019.03.162.

\bibitem{QinIJHMT19} Qin J, Luo W, Yang P, Wang B, Deng T, Han TC. Experimental demonstration of irregular thermal carpet cloaks with natural bulk material. Int J Heat Mass Transf 2019; 141: 487-490. https://doi.org/10.1016/j.ijheatmasstransfer.2019.06.092.





\bibitem{PREDai} Dai G, Shang J, Huang JP. Theory of transformation thermal convection for creeping flow in porous media: Cloaking, concentrating, and camouflage. Phys Rev E 2018; 97(2): 022129. https://doi.org/10.1103/PhysRevE.97.022129.

\bibitem{HJ-PRE2018} Xu LJ, Yang S, Huang JP. Thermal theory for heterogeneously architected structure: Fundamentals and application. Phys Rev E 2018; 98(5): 052128. https://doi:10.1103/PhysRevE.98.052128.

\bibitem{HJ-JAP2018} Dai GL, Huang JP. A transient regime for transforming thermal convection: Cloaking, concentrating, and rotating creeping flow and heat flux. J Appl Phys 2018; 124(23): 235103. https://doi.org/10.1063/1.5051524.

\bibitem{AFM2020} Peng YG, Li Y, Cao PC, Zhu XF, Qiu CW. 3D Printed Meta-Helmet for Wide-Angle Thermal Camouflages. Adv Funct Mater 2020; 30(28): 2002061. https://doi.org/10.1002/adfm.202002061.

\bibitem{HJ-PRAP2020} Xu LJ, Dai GL, Huang JP. Transformation Multithermotics: Controlling Radiation and Conduction Simultaneously. Phys Rev Appl 2020; 13(2): 024063. https://doi:10.1103/PhysRevApplied.13.024063.

\bibitem{HJ-ES2020} Xu LJ, Yang S, Dai GL, Huang JP. Transformation Omnithermotics: Simultaneous Manipulation of Three Basic Modes of Heat Transfer. ES energy environ 2020; 7: 65-70. https://doi.org/10.30919/esee8c372.

\bibitem{AM2021} Xu L, Chen H. Transformation Metamaterials. Adv Mater 2021; 33(52): 2005489. https://doi.org/10.1002/adma.202005489.

\bibitem{NRM2021} Li Y, Li W, Han TC, Zheng X, Li JX, Li BW, et al. Transforming heat transfer with thermal metamaterials and devices. Nat Rev Mater 2021; 6: 488-507. https://doi.org/10.1038/s41578-021-00283-2. 




\bibitem{PRL2014.1} Xu HY, Shi XH, Gao F, Sun HD, Zhang BL. Ultrathin three-dimensional thermal cloak. Phys Rev Lett 2014; 112(5): 054301. https://doi.org/10.1103/PhysRevLett.112.054301.

\bibitem{PRL2014.2} Han TC, Bai X, Gao DL, Thong JTL, Li BW, Qiu CW. Experimental demonstration of a bilayer thermal cloak. Phys Rev Lett 2014; 112(5): 054302. https://doi.org/10.1103/PhysRevLett.112.054302.

\bibitem{PRL2014.3} Ma YG, Liu YC, Raza M, Wang YD, He SL. Experimental Demonstration of a Multiphysics Cloak: Manipulating Heat Flux and Electric Current Simultaneously. Phys Rev Lett 2014; 113(20): 205501. https://doi.org/10.1103/PhysRevLett.113.205501.

\bibitem{HanAM18} Han TC, Yang P, Li Y, Lei DY , Li BW, Hippalgaonkar K, et al. Full-Parameter Omnidirectional Thermal Metadevices of Anisotropic Geometry. Adv Mater 2018; 30(49): 1804019. https://doi.org/10.1002/adma.201804019.

\bibitem{PRAP2022} Dai GL, Zhou YH, Wang J, Yang FB, Qu T, Huang JP. Convective Cloak in Hele-Shaw Cells with Bilayer Structures: Hiding Objects from Heat and Fluid Motion Simultaneously. Phys Rev Appl 2022; 17(4): 044006. https://doi.org/10.1103/PhysRevApplied.17.044006.





\bibitem{FujiiAPL18} Fujii G, Akimoto Y, Takahashi M. Exploring optimal topology of thermal cloaks by CMA-ES. Appl Phys Lett 2018; 112(6): 061108. https://doi.org/10.1063/1.5016090.

\bibitem{ShaNC21} Sha W, Xiao M, Zhang JH, Ren XC, Zhu Z, Zhang Y, et al. Robustly printable freeform thermal metamaterials. Nat Commun 2021; 12: 7228. https://doi.org/10.1038/s41467-021-27543-7.

\bibitem{JiIJHMT22} Ji QX, Qi YC, Liu CW, Meng SH, Liang J, Kadic M, et al. Deep learning based design of thermal metadevices. Int J Heat Mass Transf 2022; 196: 123149. https://doi.org/10.1016/j.ijheatmasstransfer.2022.123149.

\bibitem{HirIJHMT22} Hirasawa K, Nakami I, Ooinoue T, Asaoka T, Fujii G. Experimental demonstration of thermal cloaking metastructures designed by topology optimization. Int J Heat Mass Transf 2022; 194: 123093. https://doi.org/10.1016/j.ijheatmasstransfer.2022.123093.

\bibitem{2022MTP2} Sha W, Xiao M, Huang M, Gao L. Topology-optimized freeform thermal metamaterials for omnidirectionally cloaking sensors, Mater Today Phys 2022; 28: 100880. https://doi.org/10.1016/j.mtphys.2022.100880.




\bibitem{HJ-ESEE2019} Gao Y, Wang ZM, Ding D, Li WJ, Ma YG, Hao Y, et al. Novel Methods to Harness Solar Radiation for Advanced Energy Applications. ES energy environ 2019; 5: 1-7. https://10.30919/esee8c328.

\bibitem{PER2021} Wu X, Wu S, Chen X, Lin H, Forsberg E, He S. An Ultra-Compact and Reproducible Fiber Tip Michelson Interferometer for High-Temperature Sensing. Prog Electromagn Res 2021; 172: 89–99. https://doi.org/10.2528/PIER21102703.

\bibitem{E2022-1} Pendry J, Zhou J, Sun J. Metamaterials: From Engineered Materials to Engineering Materials. Engineering 2022; 17:1-2. https://doi.org/10.1016/j.eng.2022.08.001.

\bibitem{E2022-2} Lu QB, Li X, Zhang XJ, Lu MH, Chen YF. Perspective: Acoustic Metamaterials in Future Engineering. Engineering 2022; 17:22-30. https://doi.org/10.1016/j.eng.2022.04.020.

\bibitem{E2023} Xing XC, Cao Y, Tian XY, Wu L. A Thermo-Tunable Metamaterial as an Actively Controlled Broadband Absorber. Engineering 2023; 20:143-152. https://doi.org/10.1016/j.eng.2022.04.028.




\bibitem{SP2020} Imran M, Zhang L, Gain AK. Advanced thermal metamaterial design for temperature control at the cloaked region. Sci Rep 2020; 10(1): 11763. https://doi.org/10.1038/s41598-020-68481-6.





\bibitem{GaoEPL13} Gao Y, Huang JP. Unconventional thermal cloak hiding an object outside the cloak. EPL 2013; 104(4): 44001. https://doi.org/10.1209/0295-5075/104/44001.

\bibitem{JAP2019} Yang S, Xu LJ, Huang JP. Thermal magnifier and external cloak in ternary component structure. J Appl Phys 2019; 125(5): 055103. https://doi.org/10.1063/1.5083185.




\bibitem{PRL2015} Li Y, Shen XY, Wu ZH, Huang JY, Chen YX, Ni YS, et al. Temperature-Dependent Transformation Thermotics: From Switchable Thermal Cloaks to Macroscopic Thermal Diodes. Phys Rev Lett 2015; 115(19): 195503. https://doi.org/10.1103/PhysRevLett.115.195503.





\bibitem{PRAP2019} Xu LJ, Yang S, Huang JP. Passive Metashells with Adaptive Thermal Conductivities: Chameleonlike Behavior and Its Origin. Phys Rev Appl 2019; 11(5): 054071. https://doi.org/10.1103/PhysRevApplied.11.054071.

\bibitem{PRAP2020} Yang FB, Tian BY, Xu LJ, Huang JP. Experimental Demonstration of Thermal Chameleonlike Rotators with Transformation-Invariant Metamaterials. Phys Rev Appl 2020; 14(5): 054024. https://doi.org/10.1103/PhysRevApplied.14.054024.





\bibitem{NM2019} Li Y, Zhu KJ, Peng YG, Li W, Yang TZ, Xu HX, et al. Thermal meta-device in analogue of zero-index photonics. Nat Mater 2019; 18: 48-54. https://doi.org/10.1038/s41563-018-0239-6.

\bibitem{NC2020} Xu GQ, Dong KC, Li Y, Li HG, Liu KP, Li LQ, et al. Tunable analog thermal material. Nat Commun 2020; 11: 6028. https://doi.org/10.1038/s41467-020-19909-0.

\bibitem{LiAM20} Li JX, Li Y, Cao PC, Yang TZ, Zhu XF, Wang WY, et al. A Continuously Tunable Solid-Like Convective Thermal Metadevice on the Reciprocal Line. Adv Mater 2020; 32(42): 2003823. https://doi.org/10.1002/adma.202003823.






\bibitem{AM2022} Guo J, Xu GQ, Tian D, Qu Z, Qiu CW. A Real-Time Self-Adaptive Thermal Metasurface. Adv Mater 2022; 34(24)17: 2200329. https://doi.org/10.1002/adma.202201093.

\bibitem{IJHMT2023} Han TC, Nangong JY, Li Y. ITR-free thermal cloak. Int J Heat Mass Transf 2023; 203: 123779. https://doi.org/10.1016/j.ijheatmasstransfer.2022.123779.





\bibitem{SA2021} Ren W, Sun Y, Zhao DL, Aili A, Zhang S, Shi CQ, et al. High-performance wearable thermoelectric generator with self-healing, recycling, and Lego-like reconfiguring capabilities. Sci Adv 2021; 7(7): eabe0586. https://doi.org/10.1126/sciadv.abe0586.

\bibitem{LSA2022} Xiang JL, Tao ZY, Li XF, Zhao YT, He Y, Guo XH, et al. Metamaterial-enabled arbitrary on-chip spatial mode manipulation. Light Sci Appl 2022; 11: 168. https://doi.org/10.1038/s41377-022-00859-9.





\bibitem{Milton02} Milton GW. The Theory of Composite. Cambridge: Cambridge University Press; 2002.





\bibitem{HJ-AIPA2015} Zhu NQ, Shen XY, Huang JP. Converting the patterns of local heat flux via thermal illusion device. AIP Adv 2015; 5(5): 053401. https://doi.org/10.1063/1.4913994.

\bibitem{HJ-APL2016} Shen XY, Jiang CR, Li Y, Huang JP. Thermal metamaterial for convergent transfer of conductive heat with high efficiency. Appl Phys Lett 2016; 109(20): 201906. https://doi.org/10.1063/1.4967986.

\bibitem{HJ-APL2017} Yang S, Xu LJ, Wang RZ, Huang JP. Full control of heat transfer in single-particle structural materials. Appl Phys Lett 2017; 111(12): 121908. https://doi.org/10.1063/1.4994729.

\bibitem{HJ-PRAP2019} Xu LJ, Yang S, Huang JP. Thermal Transparency Induced by Periodic Interparticle Interaction. Phys Rev Appl 2019; 11(3):034056. https://doi.org/10.1103/PhysRevApplied.11.034056.

\bibitem{HJ-IJHMT2020} Jin P, Xu LJ, Jiang T, Zhang L, Huang JP. Making thermal sensors accurate and invisible with an anisotropic monolayer scheme. Int J Heat Mass Transf 2020; 163: 120437. https://doi.org/10.1016/j.ijheatmasstransfer.2020.120437.

\bibitem{HJ-PRL2022} Xu LJ, Xu GQ, Huang JP, Qiu C-W. Diffusive Fizeau Drag in Spatiotemporal Thermal Metamaterials. Phys Rev Lett 2022; 128(14): 145901.  https://doi.org/10.1103/PhysRevLett.128.145901.




\bibitem{HJ-PRE2003} Huang JP, Karttunen M, Yu KW, Dong L. Dielectrophoresis of charged colloidal suspensions. Phys Rev E 2003; 67(2): 021403. https://doi.org/10.1103/PhysRevE.67.021403.

\bibitem{HJ-JAP2004} Dong L, Huang JP, Yu KW, Gu GQ. Dielectric response of graded spherical particles of anisotropic materials. J Appl Phys 2004; 95(2): 621-624. https://doi.org/10.1063/1.1633648.

\bibitem{HJ-APL2005} Huang JP, Yu KW. Magneto-controlled nonlinear optical materials. Appl Phys Lett 2005; 86(4): 041905. https://doi.org/10.1063/1.1854719.

\bibitem{HJ-JPCC2007} Gao Y, Jian YC, Zhang LF, Huang JP. Magnetophoresis of nonmagnetic particles in ferrofluids. J Phys Chem C 2007; 111(29): 10785-10791. https://doi.org/10.1021/jp0705673.

\bibitem{HJ-PA2008} Ye C, Huang JP. Non-classical oscillator model for persistent fluctuations in stock markets. Physica A 2008; 387(5-6): 1255-1263. https://doi.org/10.1016/j.physa.2007.10.050.

\bibitem{HJ-PO2013} Liu L, Wei JR, Zhang HS, Xin JH, Huang JP. A Statistical Physics View of Pitch Fluctuations in the Classical Music from Bach to Chopin: Evidence for Scaling. PLoS One 2013; 8(3): e58710. https://doi.org/10.1371/journal.pone.0058710.

\bibitem{HJ-LPCB2015} Qiu T, Meng XW, Huang JP. Nonstraight Nanochannels Transfer Water Faster Than Straight Nanochannels. J Phys Chem B 2015; 119(4): 1496-1502. https://doi.org/10.1021/jp511262w.
\end{thebibliography}
\end{document}